\documentclass[11pt]{article}
\usepackage{apacite}
\usepackage{amssymb,amsfonts,amsmath,amsthm}
\usepackage{paralist,xspace}
\usepackage{simplemargins}
\usepackage{ulem}
\usepackage{epsfig,graphicx,wrapfig}

\def\RR{\mathbb{R}}

\def\sent{\mathrm{sent}}
\def\rec{\mathrm{rec}}
\def\map{\mathrm{MAP}}
\def\ml{\mathrm{ML}}

\newtheorem{intro-theorem}{Theorem}

\newcommand{\C}{\ensuremath{\mathcal{C}}\xspace}

\renewcommand{\v}{\mathbf}

\def\OOplus={{ {{{\mathcal{O}}}}_+}}
\def\OObar={\overline {\mathcal{O}_+}}
\def\RR{\mathbb{R}}
\def\R{\mathbb{R}}

\def\xx{\check{x}}
\def\XX{\check{X}}

\def\dstim{d_{\mathrm{stim}}}

\newcommand{\od}{\stackrel{\mbox {\tiny {def}}}{=}}

\newcommand{\supp}{\operatorname{supp}}

\DeclareMathOperator*{\argmax}{arg\,max}
\DeclareMathOperator*{\argmin}{arg\,min}

\usepackage{multicol}
\usepackage{color}
\definecolor{gold}{rgb}{0.85,.66,0}
\definecolor{cherry}{rgb}{0.9,.1,.2}
\definecolor{burgundy}{rgb}{0.8,.2,.2}
\definecolor{orangered}{rgb}{0.85,.3,0}
\definecolor{orange}{rgb}{0.85,.4,0}
\definecolor{olive}{rgb}{.45,.4,0}
\definecolor{lime}{rgb}{.6,.9,0}
\definecolor{green}{rgb}{.2,.7,0}
\definecolor{grey}{rgb}{.4,.4,.2}
\definecolor{brown}{rgb}{.4,.2,.1}
\definecolor{blue}{rgb}{0,.0, .81}

\def\carina#1{{#1}}

\title{Combinatorial neural codes from a mathematical \\coding theory perspective}
\author{Carina Curto*, Vladimir Itskov*, Katherine Morrison*, Zachary Roth*, \\ and Judy L. Walker*\\
\begin{small}* Department of Mathematics, University of Nebraska-Lincoln, Lincoln, NE 68588 \end{small}}
\date{}
\begin{document}
\maketitle

\section*{Abstract}
Shannon's seminal 1948 work gave rise to two distinct areas of research: information theory
and mathematical coding theory.  While information theory has had a strong influence
on theoretical neuroscience, ideas from mathematical coding theory have received considerably
less attention.  Here we take a new look at combinatorial neural codes from a mathematical
coding theory perspective, examining the error correction capabilities of familiar receptive field codes (RF codes).
We find, perhaps surprisingly, that the high levels of redundancy present in these codes does not support
accurate error correction, although the error-correcting performance of RF codes ``catches up'' to that
of random comparison codes when a small tolerance to error is introduced.  On the other hand, 
RF codes are good at reflecting distances between represented stimuli, while the random comparison codes are not.
We suggest that a compromise in error-correcting capability may be a necessary price to pay
for a neural code whose structure serves not only error correction, but must also reflect relationships between stimuli.

\tableofcontents

\section{Introduction}

Shannon's seminal work \cite{Shannon48} gave rise to two distinct,
though related, areas of research: {\it information theory}
\cite{CoverThomas06} and {\it mathematical coding theory}
\cite{MacWilliamsSloane83, HuffmanPless03}.  While information
theory has had a strong influence on theoretical neuroscience
\cite{Attick92, BorstTheunissen99,
Riekeetal99, QuirogaPanzeri09}, ideas central to mathematical coding theory have
received considerably less attention.  This is in large part due to
the fact that the ``neural code'' is typically regarded as a
description of the mapping, or {\em encoding map}, between stimuli and
neural responses.  Because this mapping is not in general understood,
identifying which features of neural responses carry the most
information about a stimulus is often considered to be the main goal
of neural coding theory \cite{Bialeketal91, deCharmsZador00,
  Jacobsetal09, Londonetal10}.
  In particular, information-theoretic considerations have been used to suggest 
  that encoding maps ought to maximize information and minimize the {\it redundancy} 
  of stimulus representations \cite{Attneave54, Barlow61, AdelesbergerMangan92, Attick92, Riekeetal99}, 
  although recent experiments point increasingly to high levels of redundancy in retinal and cortical codes \cite{Puchallaetal05, Luczaketal09}.

In contrast, mathematical coding theory has been primarily motivated by engineering applications, where the encoding map is always assumed to be well-known and can be chosen at will.  The primary function of a ``code'' in Shannon's original work is to allow for accurate and efficient {\it error correction} following transmission across a noisy channel.    
 ``Good codes'' do this in a highly efficient manner, so as to achieve maximal channel capacity while
allowing for arbitrarily accurate error correction.  Mathematical coding theory grew out of Shannon's challenge to design good codes,
a question largely independent of either the nature of the information being transmitted or the specifics of the encoding map.
In this perspective, redundancy is critical to the function of a code, as error correction is only possible because a code 
introduces redundancy into the representation of transmitted information \cite{MacWilliamsSloane83, HuffmanPless03}.

Given this difference in perspective, can
mathematical coding theory be useful in neuroscience?  Because of the
inherent noise and variability that is evident in neural responses, it
seems intuitive that enabling error correction should also be an
important function of neural codes \cite{Schneidman06, Hopfield08,
  SreenivasanFiete11}.  Moreover, in cases where the encoding map has
become more or less understood, as in systems that exhibit robust and
reliable receptive fields, we can begin to look beyond the encoding
map and study the features of the neural code itself.  An immediate
advantage of this new perspective is that it can help to clarify the
role of redundancy.  From the viewpoint of
information theory, it may be puzzling to observe so much redundancy
in the way neurons are representing information
\cite{Barlow61}, although the advantages of
redundancy in neural coding are gaining appreciation
\cite{Barlow01,Puchallaetal05}. Experimentally, redundancy is apparent even without an
understanding of the encoding map, from the fact that only a small
fraction of the possible patterns of neural activity are actually
observed in both stimulus-evoked and spontaneous activity \cite{Luczaketal09}.  On the other hand, it is
generally assumed that redundancy in neural responses, as in good
codes, exists primarily to allow reliable signal estimation in the
presence of noisy information transmission.  This is precisely the
kind of question that mathematical coding theory can address: {\it
  Does the redundancy apparent in neural codes enable accurate and
  efficient error correction?}

To investigate this question, we take a new look at neural coding from
a mathematical coding theory perspective, focusing on error
correction in combinatorial codes derived from neurons with idealized
receptive fields.  These codes can be thought of as binary codes, with
1s and 0s denoting neurons that are ``on'' or ``off'' in response to a
given stimulus, and thus lend themselves particularly well to
traditional coding-theoretic analyses.  Although it has been recently
argued that the entorhinal grid cell code may be very good for error
correction \cite{SreenivasanFiete11}, we show that more typical
receptive field codes (RF codes), including place field codes, perform quite
poorly as compared to random codes with matching length, sparsity, and
redundancy.  The error-correcting performance of RF codes ``catches
up,'' however, when a small tolerance to error is introduced.  This error
tolerance is measured in terms of  a metric inherited from the stimulus space,
and reflects the fact that perception of parametric stimuli is
often inexact.  We conclude that the nature of the redundancy observed
in RF codes cannot be fully explained as a mechanism to improve error
correction, since these codes are far from optimal in this regard.  On
the other hand, the structure of RF codes does allow them to naturally
encode distances between stimuli, a feature that could be beneficial
for making sense of the transmitted information within the brain.  
We suggest that a compromise in error-correcting capability may be a necessary price to pay
for a neural code whose structure serves not only error correction, but must also reflect relationships between stimuli.

\section{Combinatorial neural codes}

Given a set of neurons labelled $\{1, \dots, n\} \od [n]$, we define a {\em neural
  code} $\C \subset 2^{[n]}$ as a set of subsets of the $n$ neurons, where $2^{[n]}$ denotes
  the set of all possible subsets.  In mathematical coding
theory, a {\em binary code} is simply a set of patterns in $\{0,1\}^n$.  These
notions coincide in a natural way once we identify any element of
$\{0,1\}^n$ with its {\em support},
\[
\v{c} \in \{0,1\}^n \leftrightarrow \supp(\v{c}) \od \{i \in [n] \, | \, c_i = 1\} \in
2^{[n]},
\]
and we use the two notions
interchangeably in the sequel.  The elements of the code are called
codewords: a {\em codeword} $\v{c} \in \C$ corresponds to a subset of
neurons, and serves to represent a stimulus.  Because we discard the
details of the precise timing and/or rate of neural activity, what we
mean by ``neural code'' is often referred to in the neural coding
literature as a {\em combinatorial code} \cite{OsborneBialek08}.

We will consider {\it parameters} of neural codes, such as size,
length, sparsity and redundancy.  The {\it size} of a code $\C$ is simply the
total number of codewords, $|\C|$.  The {\it length} of a code $\C
\subset 2^{[n]}$ is $n$, the number of neurons.  The (Hamming) {\it
  weight} $w_H(\v{c})$ of a codeword $\v{c} \in \C$ is the number of neurons in $\v{c}$ 
  when viewed as a subset of $[n]$ or,
alternatively, the number of $1$s in the word when viewed as an element of
$\{0,1\}^n$.  We define the {\it sparsity} $s$ of a code as the average
proportion of 1s appearing among all codewords,
$$s = \dfrac{1}{|\C|}\sum_{\v{c} \in \C}\dfrac{w_H(\v{c})}{n}.$$

Closely related to the size of a code $\C$ is the code's {\em
  redundancy},\footnote{See \cite{Puchallaetal05} and \cite{LevyBaxter96} for related notions.}
   which quantifies the idea that typically more neurons
are used than would be necessary to encode a given set of stimuli.
Formally, we define the redundancy $\rho$ of a code $\C$ of length $n$ as
\[
\rho = 1-\dfrac{\log_2(|\C|)}{n}.
\]
For example, the redundancy
of the {\em repetition code} $\C = \{\emptyset, [n]\}$ of length $n$, consisting
only of the all-zeros word and the all-ones word,
is $\rho = \frac{n-1}{n}$; this may be interpreted as saying that all
but one of the $n$ neurons are extraneous.  At the other end of the
spectrum, the redundancy of the code $\C=2^{[n]},$ consisting of all
possible subsets of $[n],$ is $\rho = 0$.  It is clear $\rho$ takes values
between $0$ and $1$, and that any
pair of codes with matching size and length will automatically have
the same redundancy.\footnote{In the coding theory literature, the
  {\em rate} of a code $\C$ of length $n$ is given by
  {\scriptsize$\dfrac{\log_2(|\C|)}{n}$}, so that the redundancy as we
  have defined it is simply 1 minus the rate.  Because ``rate'' has a
  very different meaning in neuroscience than in coding theory, we
  will avoid this term and use the notion of redundancy instead.}

\subsection{Receptive field codes (RF codes)}
Neurons in many brain areas have activity patterns that can be
characterized by receptive fields.  Abstractly, a {\it receptive
  field} is a map $f_i:X \rightarrow \RR_{\geq 0}$ from a space of
stimuli $X$ to the average (nonnegative) firing rate of a single neuron, $i$, in
response to each stimulus.  Receptive fields are computed by
correlating neural responses to independently measured external
stimuli.  We follow a common abuse of language, where both the map and
its support (i.e., the subset of $X$ where $f_i$ takes on positive
values) are referred to as receptive fields.  {\it Convex} receptive
fields are convex subsets of $X$.  The main examples we have in mind
pertain to orientation-selective neurons and hippocampal place cells.
Orientation-selective neurons have {\it tuning curves} that reflect a
neuron's preference for a particular angle \cite{WatkinsBerkley74, Ben-Yishai1995}. Place cells are neurons that have
{\it place fields}, i.e.\ each cell has a preferred (convex) region of
the animal's physical environment where it has a high firing rate
\cite{OkeefeDostrovsky, McNaughton:2006:Nat-Rev-Neurosci}.  Both tuning curves
and place fields are examples of receptive fields.\footnote{In the
  vision literature, the term ``receptive field'' is reserved for
  subsets of the visual field; here we use the term in a more general
  sense that is applicable to any modality, as in \cite{CurtoItskov08}.}

\begin{figure}[h]\label{fig1}
\begin{center}
\includegraphics[width=4in]{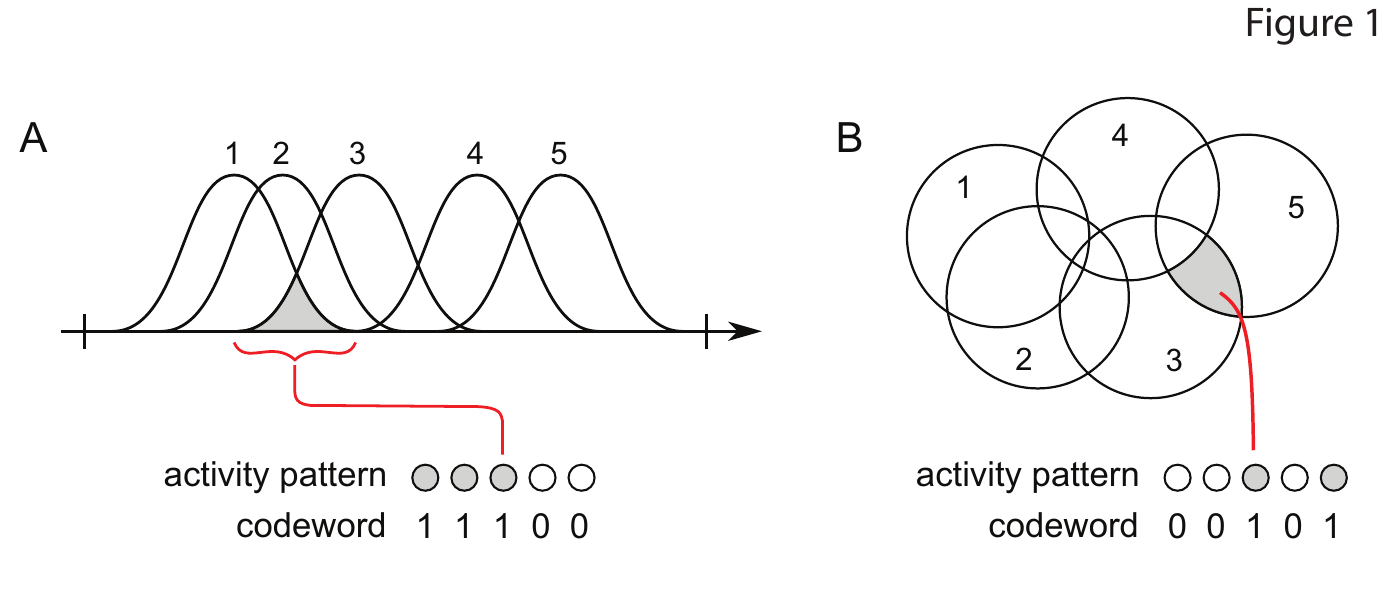}
\end{center}
\caption{\small {\bf Receptive field overlaps determine codewords in 1D and 2D RF codes.}  (A) Neurons in a 1D RF code have receptive fields that overlap on a line segment (or circle).  Each stimulus on the line
corresponds to a binary codeword, with 1s corresponding to neurons whose receptive fields include the stimulus, and 0s for neurons that are not active in response to the stimulus.  (B) Neurons in a 2D RF code, such as a place field code, have receptive fields that partition a two-dimensional region into non-overlapping intersection regions, such as the shaded area.  All stimuli within one of these regions will activate the same set of neurons, and hence have the same corresponding codeword.}
\end{figure}

The neural code is the brain's representation of the stimulus space covered by the receptive
fields.  When a stimulus lies in the intersection of several receptive
fields, the corresponding neurons tend to co-fire while the rest
remain silent.  The active subset $\sigma$ of neurons is a neural
codeword and is identified as usual with a binary codeword $\v{c}$ such that $\supp(\v{c}) = \sigma$; i.e., 
$$\v{c} = (c_1,\dots, c_n) \in \{0,1\}^n, \mbox{ where } 
c_i = \left\{\begin{array}{cc} 1, & i \in \sigma,\\ 0, & i \notin \sigma.\end{array}\right.$$
For a given set of receptive fields on a stimulus space $X$, the {\it
  receptive field code} (RF code) $\C \subset \{0,1\}^n$ is simply the
set of all binary codewords corresponding to stimuli in $X$.  The {\it
  dimension} of a RF code is the dimension of the
underlying stimulus space.\footnote{Note that this is distinct from the notion of ``dimension of a code'' in the coding theory literature.}  In the case of orientation tuning curves,
the stimulus space is the interval 
$[0,\pi)$,
and the corresponding RF code is one-dimensional.  In the case of place fields for an animal exploring a two-dimensional environment, the stimulus space is the environment, and the RF code is two-dimensional.  From now on, we will refer to such codes as 1D RF codes and 2D RF codes, respectively.

Figure 1 shows examples of receptive fields covering one- and two-dimensional stimulus spaces.  Recall that $f_i:X\rightarrow \RR_{\geq 0}$ is the receptive field of a single neuron, and let $f = (f_1,\dots, f_n):X \rightarrow \RR^n_{\geq 0}$ denote the population activity map, associating to each stimulus a firing rate vector that contains the response of each neuron as dictated by the receptive fields.  For a given choice of threshold $\theta$, we can define a {\it binary response map}, $\Phi: X \rightarrow \{0,1\}^n$, from the stimulus space $X$ to codewords by
$$\Phi_i(x) = \left\{\begin{array}{cc} 1 & \mbox{if}\;\; f_i(x)\geq \theta,\\ 0 & \mbox{if}\;\; f_i(x) < \theta. \end{array}\right.$$
 The corresponding RF code $\C$ is the image of $\Phi$.  Notice that many stimuli will produce the same binary response; in particular, $\Phi$ maps an entire region of intersecting receptive fields to the same codeword, and so $\Phi$ is far from injective. 

\subsection{Comparison codes}
In order to analyze the performance of RF codes, we will use two types
of randomly-generated comparison codes with matching size, length, and sparsity.  
In particular, these codes have the same redundancy as their corresponding RF codes.
We choose random codes as our comparison codes for three reasons.
Firstly, as demonstrated by Shannon \citeyear{Shannon48} in the proof of his channel coding
theorem, random codes are expected to have
near-optimal performance.  Secondly, the parameters can be tuned to
match those of the RF codes; we describe below the two ways in which
we do this.  Finally, random codes are a biologically reasonable
alternative for the brain, since they may be implemented by random neural networks.
\medskip

\noindent{\bf Shuffled codes.} 
   Given a RF code $\C$, we generate a {\em shuffled code} $\widetilde{\C}$
   in the following manner.  Fix a collection of permutations
   $\{\pi_{\v{c}} \, | \, \v{c} \in \C\}$ such that $(c_{\pi_\v{c}(1)},
   \dots, c_{\pi_\v{c}(n)}) \neq (c_{\pi_{\v{c}'}(1)}, \dots,
   c_{\pi_{\v{c}'}(n)})$ for all distinct $\v{c}, \v{c}' \in \C$, and set $\widetilde{\C} =
   \{(c_{\pi_\v{c}(1)}, \dots, c_{\pi_\v{c}(n)}) \mid \v{c} = (c_1,
   \dots, c_n) \in \C\}$.\footnote{If the
     same permutation were used to shuffle all codewords, the
     resulting {\em permutation equivalent} code would be nothing more
     than the code obtained from a relabelling of the neurons.}
  The shuffled code $\widetilde{\C}$ has the same
   length, size, and weight distribution (and hence the
   same sparsity and redundancy) as $\C$. 
In our simulations, each permutation $\pi_{\v{c}}$ is chosen uniformly at random with the modification that a new permutation is selected 
if the resulting shuffled codeword has already been generated.  This ensures
that no two codewords of $\C$ correspond to the same word in the shuffled code.
\medskip

\noindent {\bf Random constant-weight codes.}  {\em Constant-weight codes}
are subsets of $\{0,1\}^n$ in which all codewords have the same
weight.  Given a RF code $\C$ on $n$ neurons, we compute the average weight of the
codewords in $\C$ and round this to obtain an integer $w$.  We then
generate a constant weight code by randomly choosing
subsets of size $w$ from $[n]$.  These subsets give the positions of the codeword that are
assigned a 1, and the remaining positions are all assigned zeros.
This process is repeated until $|\C|$ distinct codewords are
generated, and the resulting code is then a random constant
weight code with the same length, size, and redundancy as $\C$, and
approximately the same sparsity as $\C$. 

\section{Stimulus encoding and decoding}

\subsection{The mathematical coding theory perspective}
The central goal of this article is to analyze our main examples of
combinatorial neural codes, 1D and 2D RF codes, from a mathematical coding theory
perspective.  We draw on this field because it provides a complementary perspective on the nature and function of codes that is unfamiliar to most neuroscientists.  We will first
discuss the standard paradigm of coding theory and then explain the
function of codes from this perspective.
Note that to put neural codes into this framework, we must discretize the stimulus space
and encoding map so that we have an injective map from the set of stimuli to the code; this will be described in the next section.

\begin{figure}[h]\label{fig2}
\begin{center}
\includegraphics[width=4.5in]{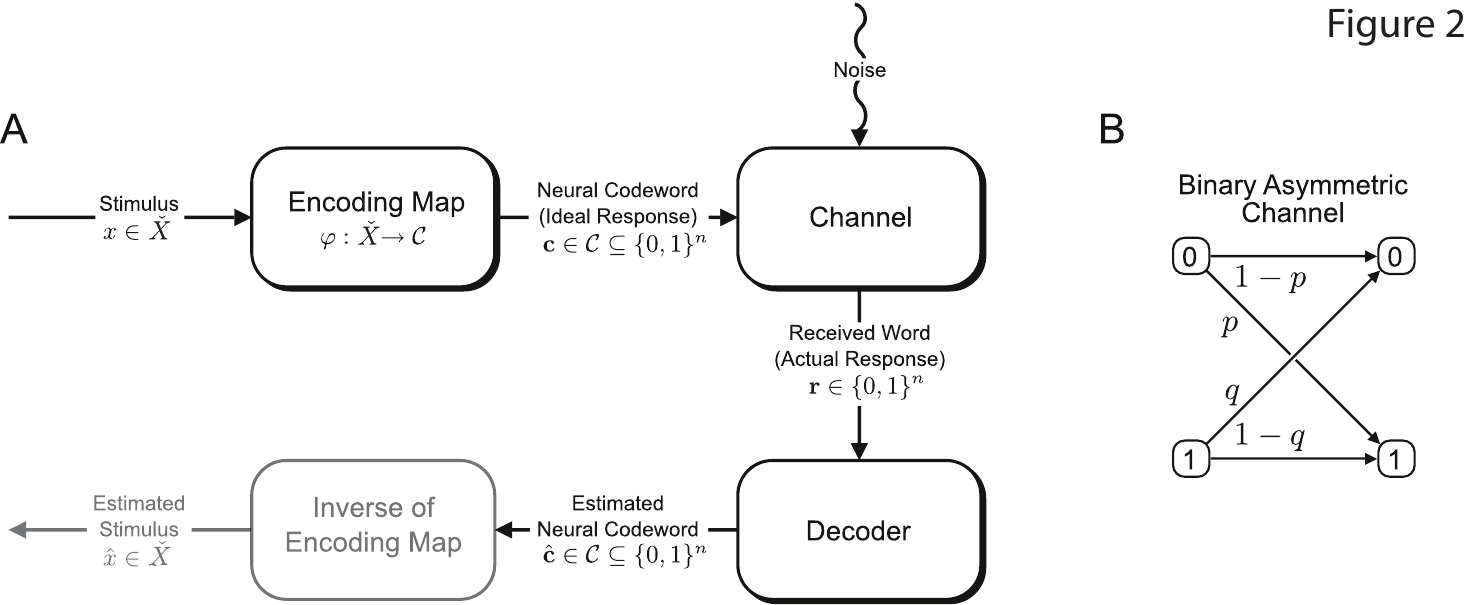}
\end{center}
\caption{\small {\bf Stimulus encoding and decoding from a mathematical coding theory perspective.}  
Here $\XX$ is the discretized stimulus space, and $\mathcal{C}$ is the neural code.  (A) A stimulus $x \in \XX$ is mapped, via an (injective) encoding map $\varphi:\XX \rightarrow \mathcal{C}$, to a neural codeword $\mathbf{c} \in \mathcal{C}$ representing the {\it ideal response} of a population of neurons.  The effect of noise is modeled by passing the codeword through a noisy channel.  The output of the channel is the received word $\mathbf{r} \in \{0,1\}^n$, representing the {\it actual response} of the population to a particular presentation of the stimulus; typically, the received word is {\it not} an element of the neural code.  In order to estimate the ideal response, the received word is passed through a decoder to produce an estimated neural codeword $\hat{\mathbf{c}} \in \mathcal{C}$.  Decoding is considered to be ``correct'' if this codeword matches the ideal response for the original stimulus.  Finally, the inverse of the encoding map can be used to identify the estimated neural codeword with an estimate for the stimulus.
(B) The Binary Asymmetric Channel (BAC) acts independently on individual bits of a binary word in $\{0,1\}^n$.  The effect of noise is to flip $0/1$ bits according to a ``false positive'' probability $p$ and a ``false negative'' probability $q$. }
\end{figure}

Figure 2A illustrates the various stages of information transmission
using the standard coding theory paradigm, adapted for RF codes.   
A stimulus $x \in \XX$ 
gets mapped to a neural codeword $\v{c} \in \C$ under an (injective) {\it encoding
  map} $\varphi:\XX \rightarrow \C$, where $\XX$ is the (discretized) stimulus space.
This map sends each stimulus to a
neural activity pattern which is considered to be the {\it ideal
  response} of a population of neurons.  The codeword, viewed as a
string of $0$s and $1$s, then passes through a noisy {\it channel},
where each 0/1 bit may be flipped with some probability.  A $1\mapsto
0$ flip corresponds to a neuron in the ideal response pattern failing
to fire, while a $0 \mapsto 1$ flip corresponds to a neuron firing when
it is not supposed to.  The resulting word is not necessarily a
codeword, and is referred to as the {\it received word}.  This noisy
channel output is then passed through a {\it decoder} to yield an
estimate $\hat{\v{c}} \in \C$ for the original codeword $\v{c}$, corresponding
to an estimate of the ideal response.  
Finally, if an estimate $\hat{x} \in \XX$ of the original stimulus is desired, the inverse of the encoding map may be applied to the estimated codeword $\hat{\v{c}}$.
Because the brain only has access to neural activity patterns, we will consider the ideal response as a proxy
for the stimulus itself;
the estimated neural codeword thus represents the brain's estimate of the stimulus, and so we can ignore this last step.

The mathematical coding theory perspective on stimulus encoding/decoding has several important differences 
from the way neuroscientists typically think about neural coding.
Firstly, there is a clear distinction made between a {\it code}, which
is simply a set of codewords (or neural response patterns) devoid of
any intrinsic `meaning,' and the {\it encoding map}, which is a
function that assigns a codeword to each element in the set of objects
to be encoded.  Secondly, this map is always 
deterministic,\carina{\footnote{\carina{In engineering applications, one can always assume the
encoding map is deterministic.  In the neuroscience context, however, it may
be equally appropriate to use a probabilistic encoding map.}}}
as the effects of noise are considered to arise purely from the transmission
of codewords through a noisy channel.  For neuroscientists, the
encoding of a signal into a pattern of neural activity is itself a
noisy process, and so the encoding map and the channel are difficult
to separate.  If we consider the output of the encoding map to be the
ideal response of a population of neurons, however, it is clear that
actual response patterns in the brain correspond not to codewords
 but rather to received words.  \carina{(The ideal response, on the
other hand, is always a codeword and corresponds intuitively to the average response across
many trials of the same stimulus.)}
In the case of
RF codes, there is a natural encoding map that sends each
stimulus to the codeword corresponding to the subset of neurons that
contain the stimulus in their receptive fields.  
In the case of the random comparison codes, an encoding map that
assigns codewords to stimuli is chosen randomly (details are given in the
next section).

Another important difference offered by the coding theory perspective is in the process of decoding.  Given a received word, the objective of the decoder is to estimate the original codeword that was transmitted through the channel.  In the case of neural codes, this amounts to taking the {\it actual} neural response and producing an estimate of the {\it ideal} response, which serves as a proxy for the stimulus.  The function of the decoder is therefore to {\it correct errors} made by transmission through the noisy channel.  In a network of neurons, this would be accomplished by network interactions that evolve the original neural response (the received word) to a closely related activity pattern (the estimated codeword) that corresponds to an ideal response for a likely stimulus.

This leads us to the coding theory perspective on the 
{\it function} (or purpose)
of a code.  Error correction is only possible when errors produced by
the channel lead to received words that are {\it not} themselves
codewords, and it is most effective when codewords are ``far away''
from each other in the space of all words, so that errors can be
corrected by returning the ``nearest'' codeword to the received word.  The
function of a code, therefore, is to represent information in a way
that allows accurate error correction in a high percentage of trials.
The fact that there is redundancy in how a code represents information
is therefore a positive feature of the code, rather than an inefficiency, since it is precisely this redundancy that makes error correction possible.

\subsection{Encoding maps and the discretization of the stimulus space}

In the definition of RF codes above, the stimulus space $X$ is a
subset of Euclidean space, having a continuum of stimuli.  Via the associated binary response maps, a set of $n$ receptive fields
 partitions the stimulus space $X$ into distinct
{\it overlap regions}, such as the shaded regions in Figure 1.  For
each codeword $\v{c} \in \C$, there is a corresponding overlap region $\Phi^{-1}(\v{c})$, all of whose points map to $\v{c}$.  The combinatorial code $\C$ therefore has
limited resolution, and is not able to distinguish between stimuli in
the same overlap region.  This leads to a natural discretization of
the stimulus space, where we assign a single representative stimulus
-- the center of mass\footnote{Although many of the overlap regions will be
  non-convex, instances of the center of mass falling outside the
  corresponding region will be rare enough that this pathological case
  need not be considered.} -- to each overlap region, and we write
$$\xx(\v{c}) \od \dfrac{\int_{\Phi^{-1}(\v{c})} x \; {dx}}{\int_{\Phi^{-1}(\v{c})} dx},$$
where $x \in X$ refers to a one or two-dimensional vector, and the integral is either a single or double integral, depending on the context.  In practice, for 2D RF codes we use a fine grid 
to determine the center of mass associated to each codeword (see Appendix B.2).

From now on, we will use the term ``stimulus space'' to refer to the {\it discretized stimulus space}:
$$\XX = \{ \xx(\v{c}) \mid \v{c} \in \C\} \subset X.$$
Note that $|\XX| = |\C|$, so we now have a one-to-one correspondence between stimuli and codewords.  The restriction of the binary response map $\Phi$ to the discretized stimulus space is the {\it encoding map} of the RF code,
$$\varphi = \Phi|_{\XX}: \XX \rightarrow \C.$$
Note that, unlike $\Phi$, the encoding map $\varphi$ is injective, and
so its inverse is well-defined.
This further supports the idea, introduced in the previous section, that
the ideal response estimate returned by the decoder can serve as a proxy
for the stimulus itself.

In the case of the comparison codes, we use the same discretized stimulus space $\XX$ as in the corresponding RF code, and associate a codeword to each stimulus using a random (one-to-one) encoding map $\varphi:\XX \rightarrow \C$.  This map is generated by ordering both the stimuli in $\XX$ and the codewords in the random code $\C$, and then selecting a random permutation to assign a codeword to each stimulus. 

\subsection{The Binary Asymmetric Channel}\label{sec:BAC}

 In all our simulations, we model the channel as a {\em binary
 asymmetric channel (BAC)}.  As seen in Figure~2B, the BAC is defined
 by a {\em false positive} probability $p$, the probability of a 0 being flipped to a 1, and a {\em false negative} probability $q$, the
 probability of a 1 being flipped to a 0.   Since errors are always
 assumed to be less likely than faithful transmission, we assume $p,q < 1/2$.
The channel operates on each individual bit, but it is customary to
extend it to operate on full codewords via the assumption that each
bit is affected independently.  This is reasonable in our context
because it is often assumed (though not necessarily believed) that
neurons within the same area experience independent and identically
distributed noise.  
The BAC has as special cases two other channels
commonly considered in mathematical coding theory: $p = q$ gives the {\em binary
  symmetric channel (BSC)}, while $p=0$ reduces to the {\em Z-channel}.

We will assume $p \leq q$, meaning that it is at least as likely that
a 1 will flip to a 0 as it is that a 0 will flip to a 1.   This is because the failure of a neuron to fire (due to, for example, synaptic failure) is considered more likely than a neuron firing when it is not supposed to.  Recall that the sparsity $s$ reflects the probability that a neuron
fires due to error-free transmission.  We will require $p < s$, as a false positive response should be less likely than a neuron firing appropriately in response to a stimulus.
Finally, since our
neural codes are assumed to be sparse, we require $s<1/2$.  In
summary, we assume:
$$p \leq q < 1/2, \;\;\text{   and   }\;\; p< s < 1/2.$$

Note that the probability of an error across this channel depends on the sparsity of the code.  For a given bit (or neuron), the probability of an error occurring during transmission across the BAC is $p(1-s) + qs,$ assuming that all codewords are transmitted with equal probability and all neurons participate in approximately the same number of codewords.

\subsection{The ML and MAP decoders}\label{MLdecoding}
A {\it decoder} takes an actual response (or received word) $\v{r} \in
\{0,1\}^n$ and returns a codeword $\hat{\v{c}}\in \C$ that is an
estimate of the ideal response (or sent word), $\v{c} \in \C$.  For
each combination of code and channel, the decoder that is {\it
  optimal}, in the sense of minimizing errors, is the one that returns
a codeword $\hat{\v{c}}$ with maximal probability\footnote{In all of
  our decoders, we assume that ties are broken randomly, with uniform
  distribution on equally-optimal codewords.} of having been sent,
given that $\v{r}$ was received.  This is called the {\it maximum a
  posteriori} (MAP) decoder, also known in the neuroscience literature as {\it Bayesian inference}
\cite{Maetal06} or an {\it ideal observer decoder}
\cite{Deneveetal99}:
$$\hat{\v{c}}_\map = \argmax_{\v{c} \in \C} P(\sent = \v{c}\mid\rec = \v{r}).$$
  Although always optimal, this decoder can be difficult to implement in
  the neural context, as it requires knowing the probabilities $P(\sent = \v{c})$ for each codeword, which is equivalent to knowing the probability distribution of stimuli.  
  
The {\it maximum likelihood} (ML) decoder
\[
\hat{\v{c}}_\ml = \argmax_{\v{c} \in \C}P(\rec = \v{r}\mid\sent =
\v{c})
\]
is much more easily implemented.  ML decoding is often used
in lieu of MAP decoding because it amounts to optimizing a
simple function that can be computed directly from the channel
parameters.  As shown in the Appendix A.1,  on the BAC with parameters $p$ and $q$ we have
\[\hat{\v{c}}_\ml = \argmax_{\v{c} \in \C} \left[ (\v{c}\cdot\v{r}) \ln
  \left(\dfrac{(1-p)(1-q)}{pq}\right)-w_H(\v{c})
  \ln\left(\dfrac{1-p}{q}\right)\right].
\]
The ML decoder thus returns
a codeword $\v{c}$ that maximizes the dot product $\v{c}\cdot\v{r} $
with the received word $\v{r}$, subject to a penalty term proportional
to its weight $w_H(\v{c})$.  In other words, it returns the codeword
that maximizes the number of matching $1$s with $\v{r}$, while minimizing the introduction of additional $1$s.

 For $p=q < 1/2$, as on the BSC, the maximization becomes (see Appendix A.1),
    \begin{eqnarray*}
     \hat{\v{c}}_\ml &=& \argmin_{\v{c} \in \C} d_H(\v{c}, \v{r}),
    \end{eqnarray*}
   where 
   $d_H(\v{c},\v{r}) = |\{i \in [n] \mid c_i \neq r_i\}|$
   is the {\it Hamming distance} between two words in $\{0,1\}^n$.  This is the well-known result that ML decoding is equivalent to Nearest Neighbor decoding, with respect to Hamming distance, on the BSC.

\begin{figure}[!ht]\label{fig3}
\begin{center}
\includegraphics[width=5in]{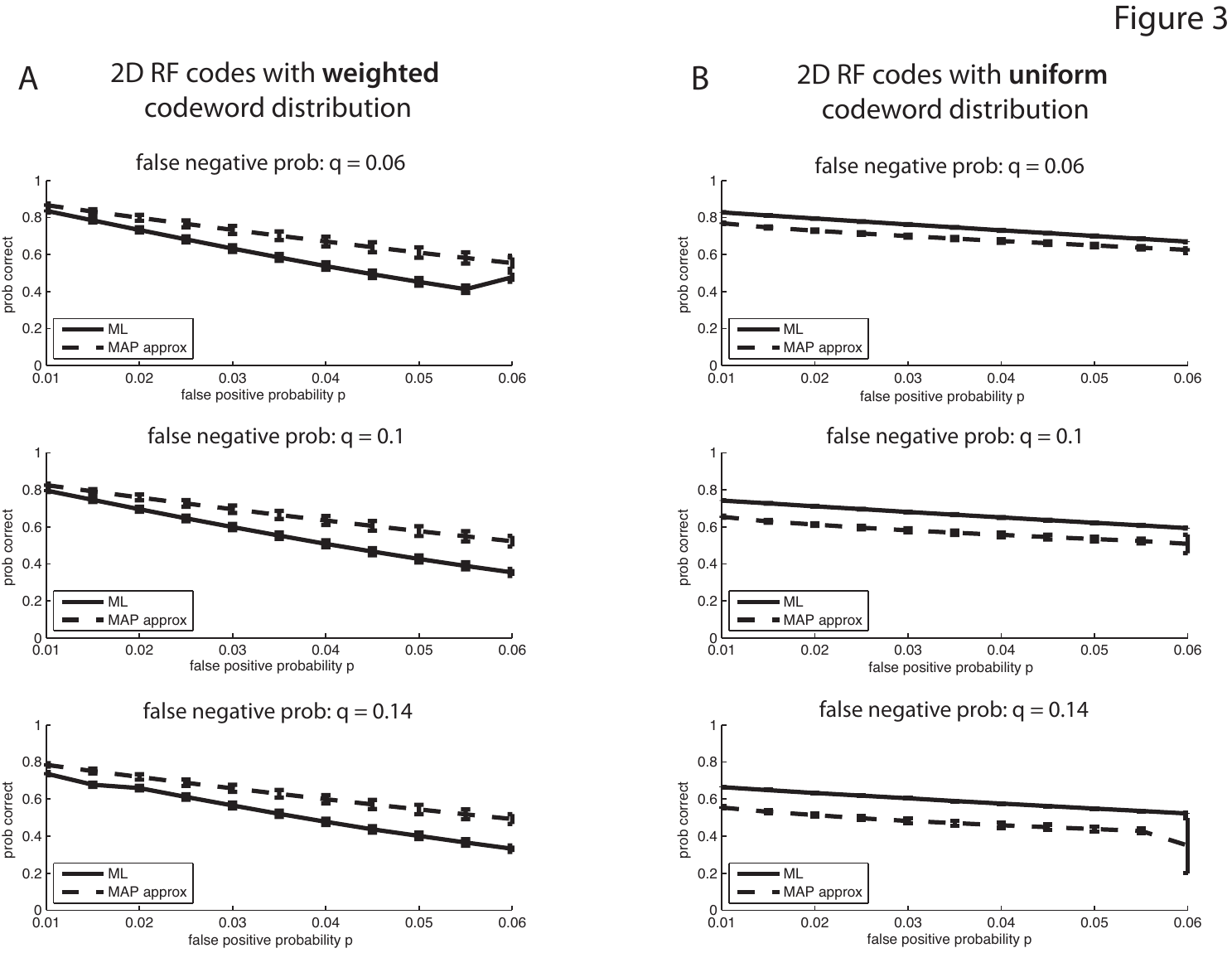}
\end{center}
\caption{\small {\bf Approximate MAP decoding outperforms ML decoding for a weighted distribution of codewords.}
(A) With the false negative probability $q$ fixed at $q=0.06$ (top), $q=0.1$ (middle) and $q=0.14$ (bottom), the false positive probability $p$ was varied in increments of $0.005$ from $0.01$ to $0.06$ to produce different channel conditions for the BAC.  On each channel, the performance of 100 2D RF codes of length 75 and mean sparsity $s=0.069$ was assessed using both the standard ML decoder and our approximation to the MAP decoder. For each BAC condition and each code, 10,000 codewords were selected according to a weighted probability distribution, where the probability of sending codeword $c$ was proportional to $s^{w_H(c)}(1-s)^{n-w_H(c)}$, as would be expected if $1$s and $0$s were sent through the channel with independent probabilities dictated by the sparsity.  
The fraction of correctly decoded words was then averaged across the 100 codes, with error bars denoting standard deviations.  The MAP approximation consistently outperformed ML decoding for all channel conditions.  (B) Same as in (A), but this time codewords were selected according to a uniform probability distribution, with each codeword equally likely.  In this case, ML decoding is equivalent to exact MAP decoding, which is always optimal.  As expected, ML decoding outperformed approximate MAP decoding for each channel condition.  Note that the error bars for ML decoding in this case are extremely small.}
\end{figure}

\subsection{An approximation of MAP decoding for sparse codes}

In cases where all codewords are sent with equal probability, it is easy to see from Bayes' rule that $\hat{\v{c}}_\ml = \hat{\v{c}}_\map$ (see Appendix A.2).  When codewords are not equally likely, MAP decoding will outperform ML decoding, but is impractical in the neural context because we cannot know the exact probability distribution on stimuli.  In some cases, however, it may be possible to approximate MAP decoding, leading to a decoder that outperforms ML while being just as easy to implement.  Here we illustrate this possibility in the case of sparse codes, where sparser (lower-weight) codewords are more likely.
   
For the BAC with parameters $p$ and $q$, and a code \C with sparsity $s$, we can approximate MAP decoding as the following maximization (see Appendix A.2):
  \[ 
\hat{\v{c}}_\map \approx \argmax_{\v{c} \in \C} \left[(\v{c}\cdot\v{r}) \ln \left(\dfrac{(1-p)(1-q)}{pq}\right)
-w_H(\v{c}) \ln\left(\dfrac{(1-p)(1-s)}{qs}\right)\right].
\] 
Since we assume that $s < 1/2$, we see that the difference between this $\hat{\v{c}}_\map$ approximation and $\hat{\v{c}}_\ml$ is only that the coefficient of the $-w_H(\v{c})$ penalty term is larger, and now depends on $s$.  Clearly, this decoder is no more difficult to implement than the ML decoder.

Figure 3 shows the results of two simulations comparing the above MAP
approximation to ML decoding on a 2D RF code.  In the first case
(Fig. 3A), the probability distribution is biased towards sparser
codewords, corresponding to stimuli covered by fewer receptive fields.  Here we see that the MAP approximation significantly outperforms ML decoding.  In the second case (Fig. 3B), all codewords are equally likely.  As expected, ML decoding outperforms the MAP approximation in this case, since it coincides with MAP decoding.  When we consider a biologically plausible probability distribution that is biased towards codewords with larger regions $\Phi^{-1}(\v{c})$ in the stimulus space, we find that ML decoding again outperforms the MAP approximation (see Appendix A.2 and Figure 8), even though there is a significant correlation between larger region size and sparser codewords.  Thus, we will restrict ourselves to considering ML decoding in the sequel; for simplicity, we will assume all codewords are equally likely.\footnote{
\carina{In cases where the distribution of stimuli is not uniform, our analysis would proceed in exactly the same manner with one exception:
instead of using the ML decoder, which may no longer be optimal, we would use the MAP decoder or an appropriate approximation to MAP
that is tailored to the characteristics of the codeword distribution.}}

\section{The role of redundancy in RF codes}

As previously mentioned, the function of a code from the mathematical
coding theory perspective is to represent information in a way that
allows errors in transmission to be corrected with high probability.
In classical mathematical coding theory, decoding
reduces to finding the closest codeword to the received word, where ``closest" is measured by a metric appropriate to the channel.  If the
code has large {\it minimum distance} between codewords, then many errors
can occur without affecting which codeword will be chosen by the
decoder \cite{HuffmanPless03}.   If, on the other hand, the elements of a binary code are closely spaced within $\{0,1\}^n$, errors will be more difficult to decode because there will often be many candidate codewords that could have reasonably resulted in a given received word.  

When the redundancy of a code is high, the ratio of the number of
codewords to the total number of vectors in $\{0,1\}^n$ is low, and so
it is possible to achieve a large minimum distance between codewords.
Nevertheless, high redundancy of a code does not guarantee large minimum
distance, because even highly redundant codes may have codewords that are spaced closely
together.  For this reason, high redundancy does not guarantee good
error-correcting properties.  This leads us to the natural question:
{\it Does the high redundancy of RF codes result in effective error
correction?}  The answer depends, of course, to some extent on the particular
decoder that is used.  In the simulations that follow, we use ML
decoding to test how well RF codes correct errors.  We assume that all
codewords within a code are equally likely, and hence ML decoding is
equivalent to (optimal) MAP decoding. It has been
sugested that the brain may actually implement ML
or MAP decoding \cite{Deneveetal99, Maetal06}, but even if this
decoder were not biologically plausible, it is the natural
decoder to use in our simulations as it provides an upper bound on the
error-correcting performance of RF codes.

\subsection{RF code redundancy does not yield effective error correction}
To test the hypothesis that the redundancy of RF codes enables effective error correction, 
we generated 1D and 2D RF codes having 75 neurons each (see
Appendix~\ref{methods}).  For each RF code, we
also generated two random comparison codes: a shuffled code and a
random constant-weight code with matching parameters.  These codes
were tested on the BAC for a variety of channel parameters (values of
$p$ and $q$).  For each BAC condition and each code, 10,000 codewords
selected uniformly at random were sent across the noisy channel and
then decoded using ML decoding.  If the decoded word exactly matched
the original sent word, the decoding was considered ``correct''; if
not, there was a failure of error-correction.

\begin{figure}[h]\label{fig4}
\begin{center}
\includegraphics[width=5in]{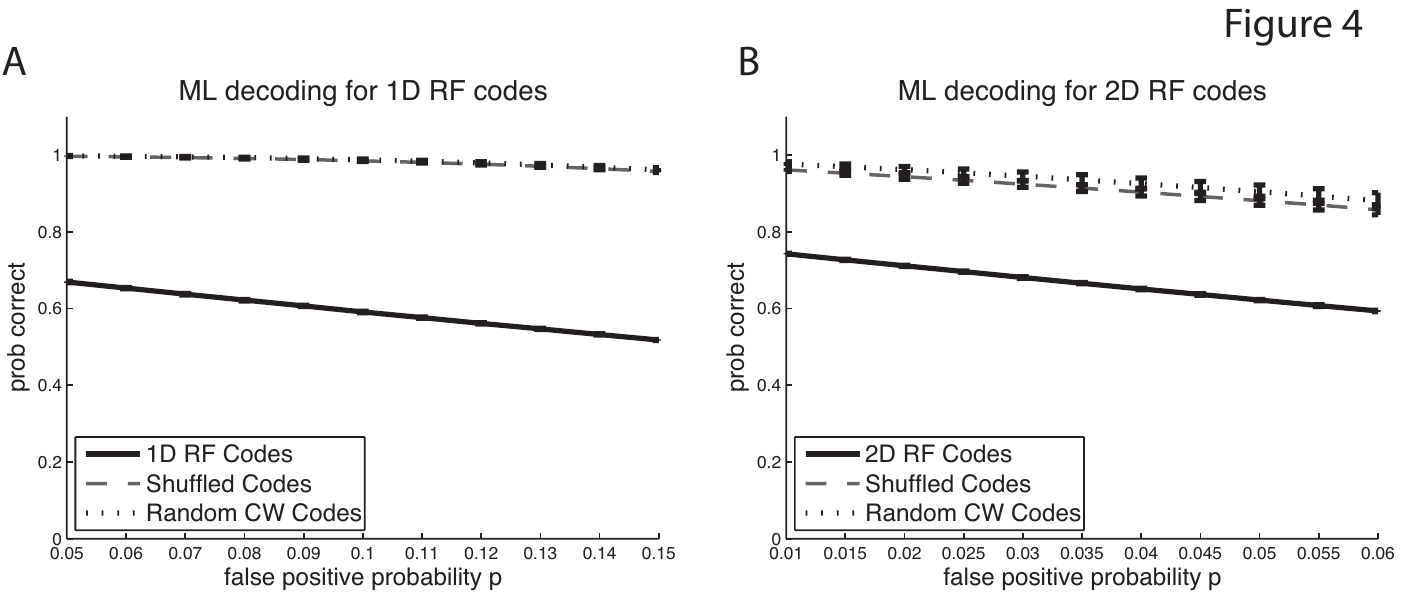}
\end{center}
\caption{\small {\bf RF codes perform poorly under standard ML decoding.} (A) With the false negative probability $q=0.2$ fixed, the false positive probability $p$ was varied in increments of $0.01$ from $0.05$ to $0.15$ to produce different channel conditions for the BAC.  On each channel, the performance of 100 1D RF codes of length 75, with mean sparsity $s=0.165$, was compared to the performance of 100 shuffled codes and 100 random constant weight codes of matched parameters.  For each BAC condition and each code, 10,000 codewords selected uniformly at random were transmitted across the BAC and then decoded with ML decoding.  The trajectories show the average performance of each code type (across the 100 sample codes) in terms of the proportion of received words that were correctly decoded.  Error bars show the magnitude of one standard deviation from the average performance, and are very small.  
While the shuffled and random constant weight codes had similar, near-optimal performance, the 1D RF codes performed quite poorly in comparison.
(B) Same as in (A), but for 2D RF codes of length 75 and mean sparsity $s=0.069$.  Here $q$ was fixed at $0.1$, while $p$ varied from $0.01$ to $0.06$ in increments of $0.005$.  Again, RF codes performed significantly worse than the shuffled and random codes with matched parameters.}
\end{figure}

Figure 4 shows the fraction of
correctly-decoded transmissions for fixed values of $q$ and a range of
$p$ values in the case of 1D RF codes (Fig. 4A) and 2D RF codes (Fig. 4B), together with the performance of the comparison codes.  In each
case, the RF codes had significantly worse performance ($< 80\%$
correct decoding in all cases) than the comparison codes, whose performances were near-optimal for low values of $p$.
Repeating this analysis for different values of $q$ yielded similar results (not shown).

As previously mentioned, in the case of the BSC,
Nearest Neighbor decoding with respect to Hamming distance
coincides with ML decoding.  Thus, in the case of a symmetric channel, codes perform
poorly precisely when their minimum Hamming distance is small.  Even
though Nearest Neighbor decoding with respect to Hamming distance
does {\it not} coincide with ML decoding on the BAC when $p \neq q$,
decoding errors are still more likely to occur if codewords are close
together in Hamming distance.  Indeed, the poor performance of RF
codes can be attributed to the very small distance between a codeword
and its nearest neighbors.  Since codewords correspond to regions
defined by overlapping receptive fields, the Hamming distance between
a codeword and its nearest neighbor is typically 1 in a RF code, which is the
worst-case-scenario.\footnote{Note that this situation would be equally
problematic if we considered the full firing rate information, instead of
a combinatorial code.  This is because small changes in firing rates
would tend to produce equally valid codewords, making error detection 
and correction just as difficult.}
In contrast, codewords in the random comparison codes
are distributed much more evenly throughout the ambient space
$\{0,1\}^n$.  While there is no guarantee that the {\em minimum}
distance on these codes is high, the {\em typical} distance between a
codeword and its nearest neighbor is high, leading to
near-optimal performance.
  
\subsection{RF code redundancy reflects the geometry of the stimulus space}
Given the poor error-correcting performance of RF codes, it seems unlikely that the primary function of RF code redundancy is to enable effective error correction.  As outlined in the previous section, the poor performance of RF
codes is the result of the very small Hamming distances between a codeword and its nearest neighbors.  While these small Hamming distances are problematic for error correction, they may prove valuable in reflecting the distance relationships between stimuli, as determined by a natural metric on the stimulus space.

To further investigate this possibility, we first define a new metric on the code that
assigns distances to pairs of codewords according to the distances between the
stimuli that they represent.  If $\v{c},\v{c'} \in \C$ are codewords,
and $\varphi:\XX \rightarrow \C$ is the (injective) encoding map, then
we define the {\it induced stimulus space metric} $\dstim:\C \times \C
\to \R_{\geq 0}$ by
\[
\dstim(\v{c},\v{c'}) = d(\varphi^{-1}(\v{c}),\varphi^{-1}(\v{c'})),
\]
where $d$ is the natural metric on the (discretized) stimulus space $\XX$.  For example,
in the case of 2D RF codes, the stimulus space is the two-dimensional
environment, and the natural metric is the Euclidean metric; in the
case of 1D RF codes, the stimulus space is $[0,\pi)$, and the natural
metric is the difference between angles, where $0$ and $\pi$ have been identified
so that, for example, $d(\pi/6,5\pi/6)=\pi/3$.

\begin{figure}[h]\label{fig5}
\begin{center}
\includegraphics[width=6in]{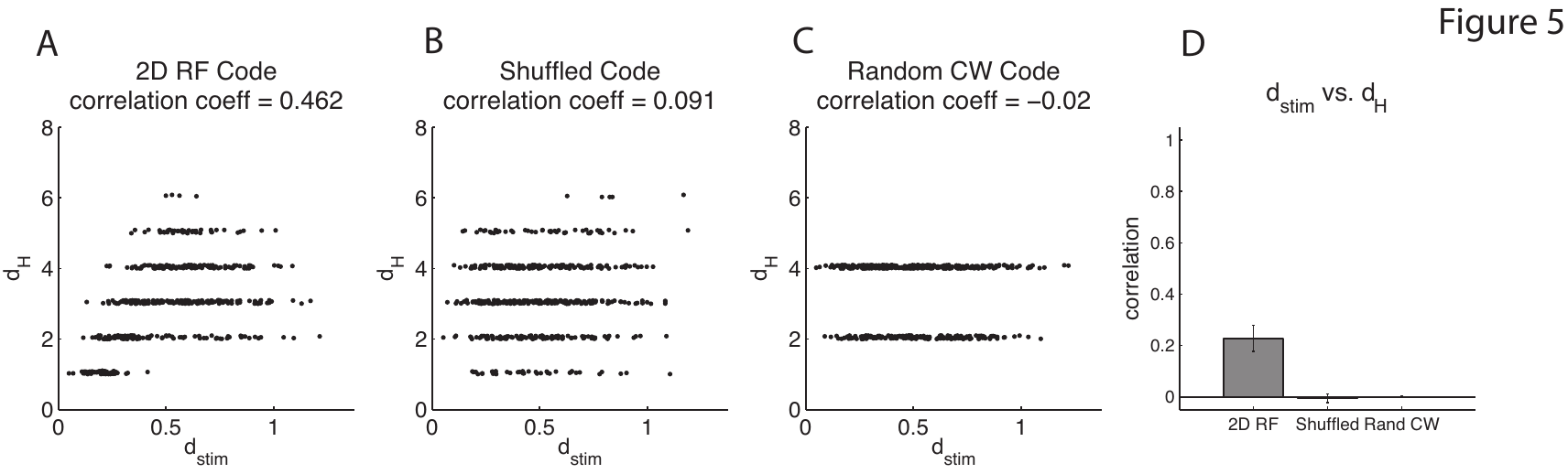}
\end{center}
\caption{\small {\bf RF codes reflect the geometry of the stimulus space.} (A) The scatter plot shows the high level of correlation (corr coeff = 0.462) between $d_{\mathrm{stim}}$ and $d_{H}$ for an example 2D RF code of length 10 and sparsity $s=0.188$.  A single code of length 10 was chosen from those generated for Figure 7 to make the correlation more visually apparent.  Each point in the scatter plot corresponds to a pair of distinct codewords.  Random noise of size at most $0.001$ in each dimension was added to each data point in order to separate points having exactly the same values of $d_H$ and $d_{\mathrm{stim}}$.  (B) Same as (A), but for a shuffled code with matched parameters.  (C) Same as in (A), but for a random constant weight code with matched parameters. (D) Average correlation coefficient between $d_{\mathrm{stim}}$ and $d_H$ for 100 2D RF codes of length 75 and mean sparsity $s=0.069$, and for 100 shuffled and random constant weight code of matched parameters (the same codes were used in Figure 4).  While the Hamming distance $d_H$ correlates significantly with stimulus space distance $d_{\mathrm{stim}}$ in the case of
RF codes, there is no correlation in the case of the random or shuffled codes.}
\end{figure}

To characterize the relationship between $\dstim$ and $d_H$ on RF codes, we
performed correlation analyses between these metrics on 2D RF
codes and corresponding random comparison codes.  For each code, we
computed $\dstim$ and $d_H$ for all pairs of codewords, and then
computed the correlation coefficient between their values. 
Figure 5A shows a scatterplot of $\dstim$ versus $d_H$ values
for a single 2D RF code; the high correlation is easily seen by eye.  In
contrast, the same analysis for a corresponding shuffled code
(Fig. 5B) and a random constant weight code (Fig. 5C) revealed no
significant correlation between $\dstim$ and $d_H$.  Repeating this
analysis for the RF and comparison codes used in Figure 4 resulted in very
similar results (Fig. 5D).  Thus, the codewords in RF codes appear to be distributed across $\{0,1\}^n$ in a
way that captures the geometry of the underlying stimulus space, rather than
in a manner that guarantees high distance between neighboring codewords.

Previous work has shown that the structure of a place field code (i.e., a 2D
RF code) can be used to extract topological and geometric features of
the represented environment \cite{CurtoItskov08}.  We hypothesize that
the primary role of RF code redundancy may be to reflect the geometry
of the underlying stimulus space, and that the poor error-correcting
performance of RF codes may be a necessary price to pay for this
feature.  This poor error correction may be mitigated, however, when
we re-examine the role that stimulus space geometry plays in the
brain's perception of parametric stimuli.

\section{Decoding with error tolerance in RF codes}

\subsection{Error tolerance based on the geometry of stimulus space}
The brain often makes errors in estimating stimuli
\cite{vanderHeijdenetal99, Prinzmetaletal01, Huttenlocheretal07}; these
errors are considered tolerable if they result in the perception of
nearby stimuli. For example,
an angle of 32 degrees might be perceived as a 30-degree angle, or a
precise position $(x,y)$ in the plane might be perceived as
$(x+\varepsilon_x, y+\varepsilon_y)$.  If the errors are relatively
small, as measured by a natural metric on the stimulus space, it is reasonable to declare the signal transmission to have
been successful, rather than incorrect.  To do this, we introduce the
notion of {\it error tolerance} into our stimulus encoding/decoding
paradigm using the induced stimulus space metric $\dstim$.
Specifically, we can decode with an error tolerance of $\delta$ by
declaring decoding to be ``correct'' if the decoded word $\hat{\v{c}}$
is within $\delta$ of the original sent word $\v{c}$:
$$\dstim(\hat{\v{c}},\v{c}) < \delta.$$
This corresponds to the perceived stimulus being within a distance $\delta$ of the actual stimulus.  

\subsection{RF codes ``catch up'' to comparison codes when decoding with error tolerance}

We next investigated whether the performance of RF codes improved, as
compared to the comparison codes with matching parameters, when an
error tolerance was introduced.  For each 1D RF code and each 2D RF
code used in Figure 4 we repeated the analysis, using fixed
channel parameters and varying instead the error tolerance with respect to the
induced stimulus space metric $\dstim$.  We found that RF codes
quickly ``catch up'' to the random comparison codes when a small
tolerance to error is introduced (Figure 6A,B).  In some cases, the performance of the
RF codes even surpasses that of the random comparison codes.
  
\begin{figure}[!h]\label{fig6}
\begin{center}
\includegraphics[width=5.5in]{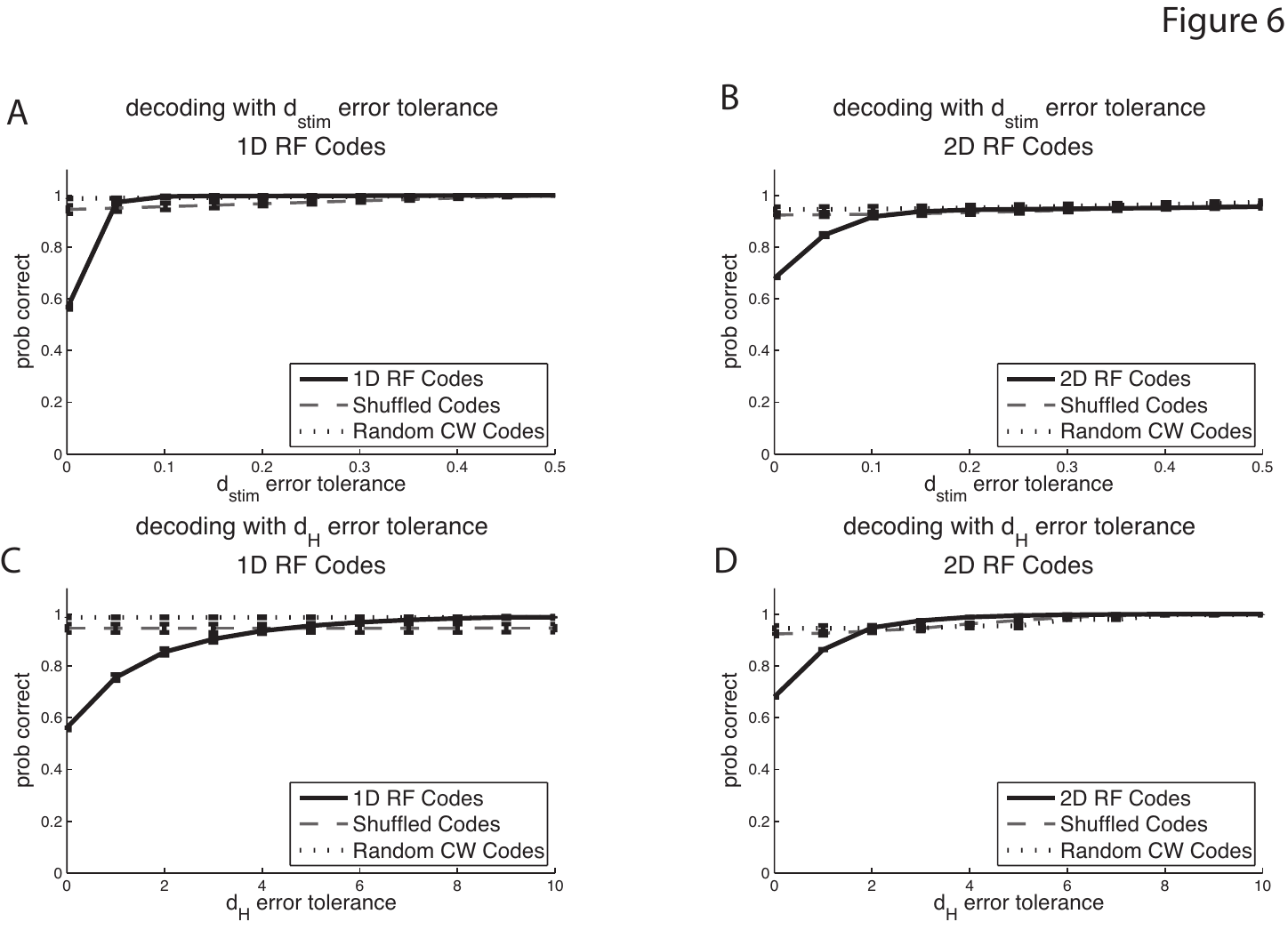}
\end{center}
\caption{\small {\bf RF codes ``catch up" in error-correcting performance when an error tolerance is introduced.}  (A) For a fixed BAC condition ($p=0.1$, $q=0.2$), the performance of 100 1D RF codes  of length 75 was compared to the performance of 100 shuffled codes and 100 random constant weight codes of matched parameters (the same codes were used in Figure 4).  For each code, 10,000 codewords selected uniformly at random were transmitted across the BAC and then decoded with ML decoding.  For each level of {\it error tolerance}, decoding was considered to be ``correct'' if the estimated word was within the given stimulus space distance, $d_{\mathrm{stim}}$, of the correct word.  
Error bars show the magnitude of one standard deviation from the average performance.  (B) Same as in (A), but for 100 2D RF codes of length 75 and corresponding shuffled and random constant weight codes of matched parameters (again, same codes as in Figure 4).  Here the channel condition was fixed at $p=0.03$ and $q=0.1$.
 (C-D) These plots are analogous to (A-B), but with the error tolerance measured using Hamming distance, $d_H$, rather than $d_{\mathrm{stim}}$.  In each case, the RF codes ``catch up'' in error-correcting performance for a small tolerance to error, at times even outperforming the shuffled and random constant weight codes.  (Note: the maximum possible value for $d_{\mathrm{stim}}$ is normalized to be $1$ in the 1D case, while in the 2D case it is $\sqrt{2}$, corresponding to the maximum distance between two points in a $1 \times 1$ square
 box enviroment.) }
\end{figure}

In order to verify that the catch-up effect is not merely an artifact
resulting from the assignment of
random encoding maps to the comparison
codes, we repeated the above analysis using Hamming distance $d_H$
instead of $\dstim$, thus completely eliminating the influence of the
encoding maps.  The Hamming distance between codewords in a sparse
code typically ranges from 0 to about twice the average weight, which
corresponds to $d_H= 25$ for the 1D RF codes, and $d_H=10$ for the
2D RF codes considered here.  Probability of correct decoding using an error
tolerance measured by Hamming distance
yielded similar results, with RF codes catching up to the random
comparison codes for relatively small error tolerances (Figure 6C,D).
This suggests that errors in transmission and decoding for RF codes
result in codewords that are close to the correct word not only in the
induced stimulus space metric, but also in Hamming distance.  

The question that remains is now: {\it Why do RF codes catch up?}

\subsection{ML similarity and ML distance}
In order to gain a better understanding of why the performance of RF
codes catches up to that of the random comparison codes when we
allow for error tolerance, we introduce the notions of {\em ML
  similarity} and {\em ML distance}\footnote{Another distance measure on neural codes was recently introduced in \cite{Tkaciketal12}.}.  Roughly speaking, the ML
similarity between two codewords $\v{a}$ and $\v{b}$ is the
probability that $\v{a}$ and $\v{b}$ will be confused in the process
of transmission across the BAC and then running the channel output
through an ML decoder. 
 More precisely, let
$\v{r}_{\v{a}}$ and $\v{r}_{\v{b}}$ be the outputs of the channel when
$\v{a}$ and $\v{b}$ are input, respectively.  Note that
$\v{r}_{\v{a}}$ (resp., $\v{r}_{\v{b}}$) is randomly chosen from
$\{0,1\}^n$, with probability distribution determined by the channel parameters
$p$ and $q$ and by the sent word $\v{a}$ (resp., $\v{b}$).  By definition, any ML
decoder will return an ML codeword given by $\argmax_{\v{c} \in \C}
P(\rec = \v{r}\, \mid\, \sent = \v{c})$ when $\v{r}$ is received from
the channel, but this ML codeword need not be unique.  To account for
this, let $\lambda(\v{r})$ be the set of ML codewords corresponding to
the received word $\v{r}$.  We then define the {\em ML similarity}\footnote{Note that this definition 
does not explicitly depend on the channel parameters, although details of the channel
are implicitly used in the computation of $P(\rec|\sent)$.}  
between the codewords $\v{a}$ and $\v{b}$ to be the probability that
the same word will be chosen (uniformly at random) from each of the
sets $\lambda(\v{r}_{\v{a}})$ and $\lambda(\v{r}_{\v{b}})$:
\[
\mu_\ml(\v{a},\v{b}) \od \sum_{\v{r}_{\v{a}}} \sum_{\v{r}_{\v{b}}} P(\rec
= \v{r}_{\v{a}} \, | \, \sent = \v{a}) P(\rec
= \v{r}_{\v{b}} \, | \, \sent = \v{b}) \frac{\left|\lambda(\v{r}_{\v{a}})
    \cap
    \lambda(\v{r}_{\v{b}})\right|}{\left|\lambda(\v{r}_{\v{a}})\right|
  \left|\lambda(\v{r}_{\v{b}})\right|}.
\]
In other words, $\mu_\ml(\v{a},\v{b})$ is the probability that if $\v{a}$ and $\v{b}$ are each sent across
the channel,
then the same codeword will be returned in each case by the decoder.  In particular, $\mu_\ml(\v{a},\v{a})$ is the probability that the same word will be returned after sending $\v{a}$ twice across the channel and decoding.  Note that typically  $\mu_\ml(\v{a},\v{a})<1$, and $\mu_\ml(\v{a},\v{a}) \neq \mu_\ml(\v{b},\v{b})$ for $\v{a} \neq \v{b}$.

In order to compare $\mu_\ml$ to distance measures such as $\dstim$ and
$d_H$, we can use the usual trick of taking the negative of the logarithm in order to convert similarity to distance:
\[
\tilde{d}_\ml(\v{a},\v{b}) \od
-\ln \mu_\ml(\v{a},\v{b}).
\]
It is clear, however, that $\tilde{d}_\ml$ is not a metric, because
$\tilde{d}_\ml(\v{a},\v{a}) \neq 0$ in general.  We can fix this
problem by first normalizing,
\[
d_\ml(\v{a},\v{b}) \od
-\ln\left(\frac{\mu_\ml(\v{a},\v{b})}{\sqrt{\mu_\ml(\v{a},\v{a})\mu_\ml(\v{b},\v{b})}}\right),
\]
so that $d_\ml(\v{a},\v{a}) = 0$ for all words in $\{0,1\}^n$.  We call $d_\ml$ the {\em ML distance}.  Unfortunately,
$d_\ml$ still fails to be a metric on $\{0,1\}^n$, as the triangle
inequality is not generally satisfied (see Appendix A.3), although it may be a metric when restricted to a particular code.

Despite not being a metric on $\{0,1\}^n$, $d_\ml$ is 
useful as an indicator of how close the ML decoder comes to outputting the correct
idealized codeword.
By definition, ML decoding errors will have large ML similarity to the
correct codeword. In other words, even if
$\hat{\v{c}}_\ml \neq \v{c}$, the value of
$d_\ml(\hat{\v{c}}_\ml,\v{c})$ will be relatively small.  
Unlike Hamming distance,
$d_\ml$ naturally captures the notion that two codewords
are  ``close'' if they are likely to be confused after having been sent
through the BAC channel and decoded with the ML decoder.\footnote{On the binary {\it symmetric} channel (BSC), Hamming distance
  \emph{does} measure the likelihood of two codewords being confused after
  ML-decoding of errors introduced by the channel.}
In practice, however, $d_\ml$
is much more difficult to compute than Hamming distance. Fortunately, as we will see in the next
section, there is a high correlation between $d_\ml$ and $d_H$, so that $d_H$ may be used
as a proxy for $d_\ml$ when using $d_\ml$ becomes computationally intractable.

\subsection{Explanation of the ``catch-up'' phenomenon}

The ML distance $d_\ml$ is defined so that ML decoding errors have small ML distance to the correct codeword, irrespective of the code.
On the other hand, tolerating small errors only makes sense if errors are quantified by distances between stimuli, given by the induced stimulus space metric $\dstim$.
The fact that RF codes catch up in error-correction when an error tolerance with respect to $\dstim$ is introduced suggests that, on these codes, $\dstim$ and $d_\ml$ correlate well, 
whereas on the comparison codes they do not.  In other words, even though the codewords in RF codes are not well-separated inside $\{0,1\}^n$, decoding errors tend to return codewords that
represent very similar stimuli, and are hence largely tolerable.

\begin{figure}[!h]\label{fig7}
\begin{center}
\includegraphics[width=5in]{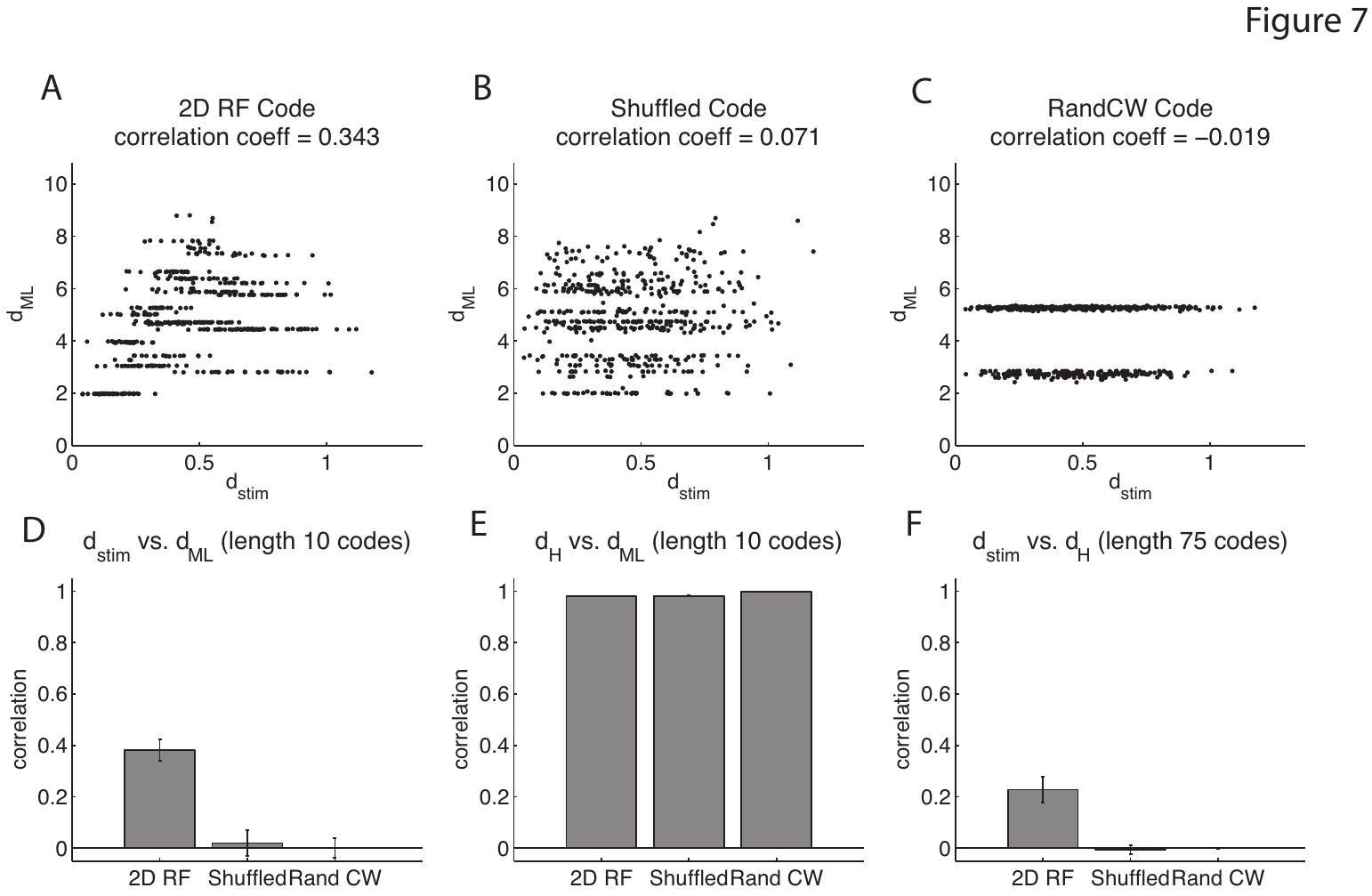}
\end{center}
\caption{\small {\bf Correlations between stimulus space distance $d_{\mathrm{stim}}$, Hamming distance $d_H$, and ML distance $d_{\mathrm{ML}}$.}  (A) The scatter plot shows the correlation between $d_{\mathrm{stim}}$ and $d_{\mathrm{ML}}$ for a single 2D RF code of length 10 and sparsity $s=0.188$.  (A code of length 10 was chosen because $d_{\mathrm{ML}}$ is computationally intractable for longer codes.) 
Each point in the scatter plot corresponds to a pair of distinct codewords;
random noise of size at most $0.001$ in each dimension was added to each data point in order to separate points having exactly the same values of $d_{\mathrm{stim}}$ and 
$d_{\mathrm{ML}}$. 
The values for $d_{\mathrm{ML}}$ were computed for channel parameters $p=0.03$ and $q=0.1$.  
(B) Same as in (A), but for a shuffled code with matched parameters to the 2D RF code.  (C) Same as in (A), but for a random constant weight code with matched parameters.
(D) Average correlation coefficient between $d_{\mathrm{stim}}$ and $d_{\mathrm{ML}}$ for ten 2D RF codes of length 10 and mean sparsity $s=0.191$, and ten shuffled and random constant weight codes of matched parameters. All $d_{\mathrm{ML}}$ values were computed for channel parameters $p=0.03$ and $q=0.1$.  Error bars denote standard deviations. (E) Average correlation coefficient between $d_H$ and $d_{\mathrm{ML}}$ for the same codes and channel condition used in (D). 
The high correlation across codes suggests that $d_H$ may be used as a proxy for $d_{\mathrm{ML}}$ in cases where $d_{\mathrm{ML}}$ is computationally intractable.
(F)  Average correlation coefficient between $d_{\mathrm{stim}}$ and $d_H$ for 100 2D RF codes of length 75 and mean sparsity $s=0.069$, and for 100 shuffled and random constant weight code of matched parameters (same codes as in Figure 4; note that this panel reproduces Figure 5D).  Here we think of $d_H$ as a proxy for $d_{\mathrm{ML}}$.
Because $d_{\mathrm{ML}}$ was not computed for this plot, calculations involving the larger codes were computationally feasible. Given the correlation patterns in (A-F), it is
likely that $d_{\mathrm{stim}}$ and $d_{\mathrm{ML}}$ are significantly correlated for large RF codes, but not for the shuffled or random constant weight comparison codes.}
\end{figure}

To verify this intuition we performed correlation analyses between
$\dstim$ and $d_\ml$ on 2D RF codes and corresponding random comparison
codes.  For each code, we computed $\dstim$ and $d_\ml$ for all pairs of
codewords, and then computed the correlation coefficient between these
two measures.  Because finding $d_\ml$ among.irs of codewords in
a code with many neurons was computationally intractable, we performed
this analysis on short codes having only 10 neurons, or length 10.
Figure 7A shows a scatterplot of $\dstim$ versus $d_\ml$ values for a
single 2D RF code; the high correlation is easily seen by eye.  In contrast, the same
analysis for a corresponding shuffled code (Fig. 7B) and a random
constant weight code (Fig. 7C) revealed no significant correlation
between $\dstim$ and $d_\ml$.  Repeating this analysis for a total of 10
matched sets of codes, each consisting of a 2D RF code, a
corresponding shuffled code, and a corresponding random constant
weight code, resulted in very similar results (Fig. 7D).

In order to test if the correlation between $\dstim$ and $d_\ml$ might continue to hold for our longer
codes with $n=75$ neurons, we first investigated whether Hamming distance $d_H$ could be
used as a proxy for $d_\ml$, as the latter can be computationally intractable.
Indeed, on all of our length 10 codes we found near perfect
correlation between $d_H$ and $d_\ml$ (Fig. 7E).  We then computed correlation
  coefficients using $d_H$ instead of $d_\ml$ for the length 75 2D
  RF codes, and corresponding comparison codes, that were analyzed in
  Figures 4, 5 and 6.  As expected, there was a significant correlation
  between $\dstim$ and $d_H$ for RF codes, but not for the random
  comparison codes (Fig. 7F).  It is thus likely that
  $\dstim$ and $d_\ml$ are well-correlated for the large RF codes
  that displayed the catch-up phenomenon (Fig. 6), but not for the comparison
  codes.

\section{Discussion}
We have seen that although RF codes are highly redundant, they do not have particularly good error-correcting capability,
performing far worse than random comparison codes of matching size, length, sparsity and redundancy.
This poor performance is perhaps not surprising when we consider the close proximity between RF codewords inside
$\{0,1\}^n$, a feature that limits the number of errors that can be corrected.  On the other hand, RF code redundancy
seems well-suited for preserving {\it relationships} between encoded stimuli, allowing these codes to reflect the geometry
of the represented stimulus space.  Interestingly, RF codes quickly ``catch up''
to the random comparison codes in error-correcting capability when a small tolerance to error is introduced.
The reason for this ``catch up'' is that errors in RF codes tend to result in nearby codewords
that represent similar stimuli, a property that is not characteristic of the random comparison codes.  
Our analysis suggests that in the context of neural codes, there may be a natural trade-off between a code's efficiency/error-correcting 
capability and its ability to reflect relationships between stimuli.  It would be interesting to investigate whether RF codes
are somehow ``optimal'' in this regard, though this is beyond the scope of this paper.  Likewise, it would be interesting to test 
the biological plausibility of the error tolerance values that are required for RF codes to catch up.  
\carina{For visual orientation discrimination in human psychophysics experiments, the perceptual
errors range from about 4$^\circ$ to 12$^\circ$ (out of 180$^\circ$) \cite{shapley2004,wehrhahm2000}; this is roughly consistent with a 5\% error tolerance, a level that resulted
in complete ``catch up'' for the 1D RF codes (Figure 6A).}

\carina{Throughout this work, we have assumed that neurons are independent.  This assumption
arose as a consequence of using the BAC as a channel model for noise, which operates on each neuron independently (see Section~\ref{sec:BAC}).
While somewhat controversial \cite{Schneidmanetal03}, there is some experimental evidence that supports the independence assumption \cite{GawneRichmond93, Nirenbergetal01}, in addition to a significant body of theoretical work that suggests that ignoring noise correlations does not significantly impact the decoding of neural population responses \cite{AbbottDayan99, AverbeckLee04, LathamNirenberg05}.   Nevertheless, it is quite possible that the error-correcting capabilities of RF codes may
increase (or decrease) if this assumption is relaxed \cite{averbeck2006}.  
It would thus be interesting to explore a similar
analysis for channel models the produce correlated noise, though this is beyond the scope of the current paper. }

\carina{We have also assumed a perfect understanding of the encoding map; however, it is possible that error-correcting capabilities
 vary significantly according to what aspect of the stimulus is being represented, similar to what has been found in information-theoretic analyses 
\cite{nemenman2008}.  Furthermore, in assessing the error-correcting properties of RF codes as compared to random comparison
codes, we used a decoder that was optimal for all codes.  If instead we used a biologically-motivated decoder, such as those suggested in 
\cite{Deneveetal99, latham-pouget-2008}, the performance of the random comparison
codes may be significantly compromised, leading to a relative improvement in error correction for RF codes.}

Mathematical coding theory has been very successful in devising codes that are optimal or nearly optimal for correcting noisy transmission
errors in a variety of engineering applications \cite{MacWilliamsSloane83, Wicker:1994:RCA:527770, HuffmanPless03}.
We believe this perspective will also become increasingly fruitful in neuroscience, as it
 provides novel and rigorous methods for analyzing neural codes in cases where the encoding map is relatively
 well-understood.  In particular, mathematical coding theory can help to clarify apparent 
paradoxes in neural coding, such as the prevalence of redundancy when it is assumed that
neural circuits should maximize information.  Finally, we believe the coding theory perspective will eventually provide
the right framework for analyzing the trade-offs 
that are inherent in codes that are specialized for information transfer and processing in the brain.

\section{Acknowledgments}
CC was supported by NSF DMS 0920845 and an Alfred P. Sloan Research Fellowship.  VI was supported by NSF DMS 0967377 and NSF DMS 1122519.
KM was supported by NSF DMS 0903517 and NSF DMS 0838463. ZR was supported by Department of Education GAANN grant P200A060126.  
JLW was supported by NSF DMS 0903517.

\appendix

\section{Appendix: ML and MAP decoding}

 \subsection{ML decoding on the BAC}
 Here we derive a simple expression for the ML decoder on the binary asymmetric channel with ``false positive'' probability $p$ and ``false negative'' probability $q$, as in Figure 2B.  Recall that the ML decoder is given by
 $$\hat{\v{c}}_\ml = \argmax_{\v{c} \in \C} P(\rec = \v{r}\mid\sent = \v{c}),$$
 where $\v{r} \in \{0,1\}^n$ is the received word, or ``actual response'' of the population of $n$ neurons, and $\C$ is the neural code.  Because the channel is assumed to act on each neuron independently, $P(\rec = \v{r}\mid\sent = \v{c})$ will only depend on the following quantities:
 \begin{align*}
 t_{00}(\v{c},\v{r}) &= \mbox{ \# of 0s that match between $\v{c}$ and $\v{r}$},\\
 t_{11}(\v{c},\v{r}) &= \mbox{ \# of 1s that match between $\v{c}$ and $\v{r}$},\\
 t_{01}(\v{c},\v{r}) &= \mbox{ \# of 0s in $\v{c}$ that correspond to 1s in $\v{r}$},\\
 t_{10}(\v{c},\v{r}) &= \mbox{ \# of 1s in $\v{c}$ that correspond to 0s in $\v{r}$}.
 \end{align*}
With this, it is straightforward to compute
\[
P(\rec = \v{r}\mid\sent = \v{c}) = (1-p)^{t_{00}(\v{c},\v{r})}p^{t_{01}(\v{c},\v{r})}(1-q)^{t_{11}(\v{c},\v{r})} q^{t_{10}(\v{c},\v{r})}.
\]
 Using the obvious identities,
 \begin{align*}
 t_{01}(\v{c},\v{r}) + t_{11}(\v{c},\v{r}) &= w_H(\v{r})\\
 t_{10}(\v{c},\v{r}) + t_{00}(\v{c},\v{r}) &= n - w_H(\v{r}),
 \end{align*}
we find
 $$P(\rec = \v{r}\mid\sent = \v{c}) = (1-p)^{t_{00}(\v{c},\v{r})}p^{w_H(\v{r})-t_{11}(\v{c},\v{r})}(1-q)^{t_{11}(\v{c},\v{r})} 
 q^{n-w_H(\v{r})-t_{00}(\v{c},\v{r})}.$$
 When we do the maximization over $\v{c}\in \C$, we can ignore terms that are independent of $\v{c}$, and we obtain
 \begin{align*}\label{BACapprox}
 \hat{\v{c}}_\ml &= \argmax_{\v{c} \in \C} \left[\left(\dfrac{1-p}{q}\right)^{t_{00}(\v{c},\v{r})} \left(\dfrac{1-q}{p}\right)^{t_{11}(\v{c},\v{r})}\right]\\
 &= \argmax_{\v{c} \in \C} \left[{t_{00}(\v{c},\v{r})}\ln\left(\dfrac{1-p}{q}\right) + {t_{11}(\v{c},\v{r})}\ln\left(\dfrac{1-q}{p}\right)\right]\tag{*}
 \end{align*}
 If we further observe that
 \begin{align*}
 t_{11}(\v{c}, \v{r}) &= \v{c}\cdot\v{r},\\ 
 t_{00}(\v{c}, \v{r}) &= (\v{1}-\v{c})\cdot (\v{1}-\v{r}) = n - w_H(\v{c}) - w_H(\v{r}) + \v{c}\cdot \v{r},
 \end{align*}
where $\v{1} \in \{0,1\}^n$ is the all-ones word, and again ignore
 terms that are independent of $\v{c}$, we obtain
 \begin{equation}\label{eq:cml}
 \hat{\v{c}}_\ml = \argmax_{\v{c} \in \C} \left[ 
 (\v{c}\cdot\v{r}) \ln \left(\dfrac{(1-p)(1-q)}{pq}\right)-w_H(\v{c}) \ln\left(\dfrac{1-p}{q}\right)\right].
 \end{equation}
 Since we assume $p,q < 1/2$, the decoder maximizes the number $\v{c}\cdot\v{r}$ of matching $1$s between the sent and received words, subject to a penalty term that is proportional to the weight (i.e.\ the number of active neurons) of the sent word.
 
 Note that for $p=q < 1/2$, as on the BSC, equation~(\ref{BACapprox}) becomes
 \[
      \hat{\v{c}}_\ml = \argmax_{\v{c} \in \C} \left[ t_{00}(\v{c}, \v{r}) + t_{11}(\v{c}, \v{r})\right] 
      =  \argmax_{\v{c} \in \C} \left[n-d_H(\v{c}, \v{r})\right] 
      = \argmin_{\v{c} \in \C} \left[d_H(\v{c}, \v{r})\right],
     \]
    where 
    $$d_H(\v{c},\v{r}) = t_{01}(\v{c},\v{r}) + t_{10}(\v{c},\v{r}) =  |\{i \in [n] \mid c_i \neq r_i\}|,$$
    is the {\it Hamming distance} between two words in $\{0,1\}^n$.  In other words, ML decoding is equivalent to Nearest Neighbor decoding, with respect to Hamming distance, on the BSC.

 \subsection{Comparison of ML and MAP decoding using Bayes' rule}

Given two events $A$ and $B$ such that the probability of $B$ is
nonzero, {\em Bayes' rule} states
\[
P(A|B) = \frac{P(B|A)P(A)}{P(B)}.
\]
We can use this theorem to relate the ML and MAP decoders:  
 \begin{align*}
 \hat{\v{c}}_\map &= \argmax_{\v{c}\in\C} P(\sent = \v{c}\mid\rec = \v{r}) 
 = \argmax_{\v{c}\in\C}\dfrac{P(\rec = \v{r} \mid \sent = \v{c}) P(\sent =\v{c})}{P(\rec = \v{r})}\\
 &=\argmax_{\v{c}\in\C}P(\rec = \v{r} \mid \sent = \v{c})P(\sent =\v{c}).
 \end{align*}
In the case that all words are sent with equal probability, i.e., $P(\sent =\v{c})$ is constant over all codewords $\v{c}\in\C$, we have
 \begin{align*}
 \hat{\v{c}}_\map  &=\argmax_{\v{c}\in\C}P(\rec = \v{r} \mid \sent = \v{c}) = \hat{\v{c}}_\ml.
 \end{align*}
Thus, the two decoders coincide when all stimuli are equally likely.
In the case where some codewords are more likely to be transmitted than
others, however, MAP and ML decoding need not coincide.  

Suppose that the probability of a codeword being sent can be {\it approximated} by
assuming individual $0$s and $1$s are transmitted with independent
probabilities consistent with the sparsity $s$ of the code:
\begin{eqnarray*}
 P(\sent = 1) &\approx& s,\\
 P(\sent = 0) &\approx& 1-s.
 \end{eqnarray*}
 Under these assumptions, we approximate
 $$P(\sent = \v{c}) \approx s^{w_H(\v{c})}(1-s)^{n-w_H(\v{c})}.$$
 Using Bayes' rule and Equation~\eqref{eq:cml}, this gives an approximation for the MAP decoder as,
 \begin{align*}
 \hat{\v{c}}_\map &\approx  \argmax_{\v{c}\in\C}\; [P(\rec = \v{r} \mid \sent = \v{c})s^{w_H(\v{c})}(1-s)^{n-w_H(\v{c})}]\\
 &=  \argmax_{\v{c}\in\C}\;\left[ \ln P(\rec = \v{r} \mid \sent = \v{c}) +w_H(\v{c})\ln s + (n-w_H(\v{c}))\ln(1-s)\right]\\
 &= \argmax_{\v{c} \in \C} \left[
 (\v{c}\cdot\v{r}) \ln \left(\dfrac{(1-p)(1-q)}{pq}\right)-w_H(\v{c}) \ln\left(\dfrac{1-p}{q}\right)  -w_H(\v{c})\ln\left(\dfrac{1-s}{s}\right)\right]\\
 &= \argmax_{\v{c} \in \C} \left[(\v{c}\cdot\v{r}) \ln \left(\dfrac{(1-p)(1-q)}{pq}\right)-w_H(\v{c}) \ln\left(\dfrac{(1-p)(1-s)}{qs}\right)\right].
 \end{align*}
 Comparing this
 approximation to equation~\eqref{eq:cml} we see that the difference
 between MAP and ML for sparse codes $(s < 1/2)$ is that the approximate MAP
 decoder has a larger penalty term associated to the weight
 $w_H(\v{c})$.  This means that the approximate MAP decoder will sometimes return
 lower-weight codewords than the ML decoder.  Unlike MAP, the ML
 decoder is completely indifferent to the code sparsity parameter $s$.

\begin{figure}[!h]\label{figSupp}
\begin{center}
\includegraphics[width=2.5in]{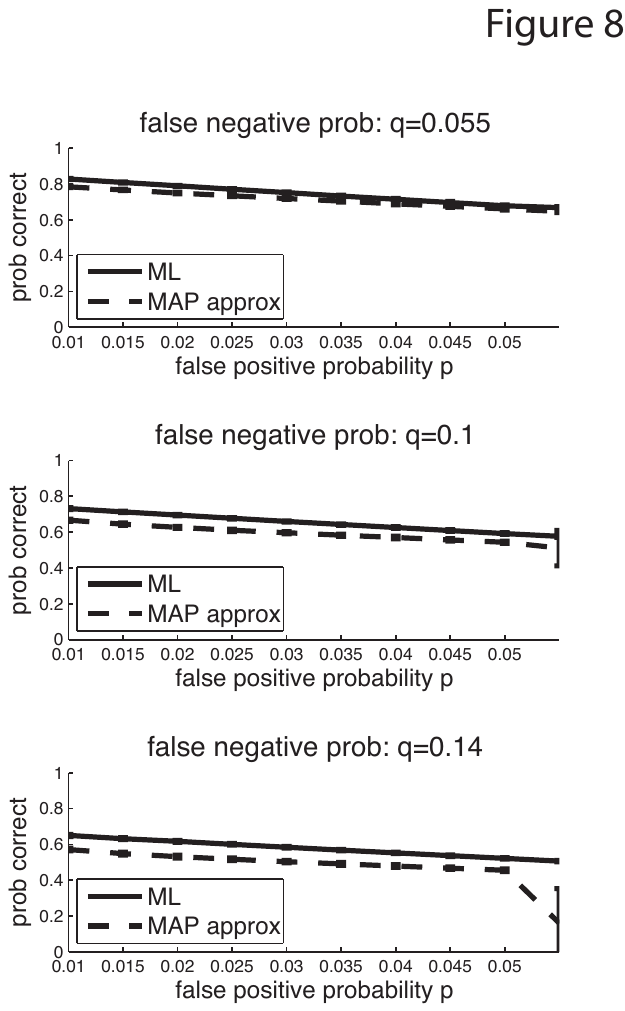}
\end{center}
\caption{\small {\bf ML decoding outperforms approximate MAP decoding for a distribution of codewords weighted by the size of the codeword region.}  With the false negative probability $q$ fixed at $q=0.055$ (top), $q=0.1$ (middle) and $q=0.14$ (bottom), the false positive probability $p$ was varied in increments of $0.005$ from $0.01$ to $0.055$ to produce different channel conditions for the BAC.  On each channel, the performance of 100 2D RF codes of length 75 and mean sparsity $s=0.596$ was assessed using the standard ML decoder and our approximation to the MAP decoder. For each BAC condition and each code, 10,000 codewords were selected according to a weighted probability distribution, where the probability of sending codeword $c$ was proportional to the area of $\varphi^{-1}(c)$, which was approximated by the fraction of points from the $300 \times 300$ fine grid that fell within the region $\varphi^{-1}(c)$ (if no grid point lay inside a region, we counted it as 1 grid point to ensure the probability was nonzero).  Each codeword was transmitted across the BAC and decoded using both the ML decoder and the approximate MAP decoder.  The fraction of correctly decoded words was then averaged across the 100 codes, with error bars denoting standard deviations.  ML decoding consistently outperformed approximate MAP decoding for each channel condition, even though the opposite was true when codewords were weighted according to the sparsity of the code (see Figure 3A).}
\end{figure}

In our simulations with 2D RF codes, we have found that the above MAP approximation outperforms ML decoding when codewords in the distribution of transmitted words are weighted in a manner dictated by the sparsity of the code (Fig. 3).  What if the codeword distribution is instead weighted by the sizes of the stimulus space regions corresponding to each codeword? In this case, Figure 8 shows that ML decoding outperforms the MAP approximation, further justifying our use of ML decoding in our analysis of the error-correcting properties of RF codes.

\subsection{Failure of the triangle inequality for $d_\ml$}
Recall that the ML distance $d_\ml$ is defined by 
\[
d_\ml(\v{a},\v{b}) \od
-\ln\left(\frac{\mu_\ml(\v{a},\v{b})}{\sqrt{\mu_\ml(\v{a},\v{a})\mu_\ml(\v{b},\v{b})}}\right),
\]
where the ML similarity $\mu_\ml$ is the probability that $\v{a}$ and
$\v{b}$ will be confused in the process of transmission across the BAC
and then running the channel output through an ML decoder.

It is clear that $d_\ml$ is a {\em pseudo-semimetric} on $\{0,1\}^n$; i.e.,
for all $\v{a}, \v{b} \in \{0,1\}^n$ we have $d_\ml(\v{a},\v{b}) \geq
0$, $d_\ml(\v{a},\v{a})=0$, and $d_\ml(\v{a},\v{b})
= d_\ml(\v{b},\v{a})$.  However, $d_\ml$ is not a {\em metric} or even
a {\em pseudo-metric} on
$\{0,1\}^n$ because it fails to satisfy the triangle inequality.  As
an example, consider the code $\C = \{(1,1,0), (1,0,1), (0,0,1)\}$,
and take $\v{x} = (0,0,1)$, $\v{y} = (0,0,0)$ and $\v{z} = (0,1,0)$.  For channel conditions $p=0.05$ and $q=0.07$,
we obtain
\[
d_\ml(\v{x},\v{y}) + d_\ml(\v{y},\v{z}) = .005 + 3.217 = 3.222 < 4.072
= d_\ml(\v{x},\v{z}).
\]
It is interesting to note, however, that both the triangle
inequality $d_\ml(\v{a},\v{b}) + d_\ml(\v{b},\v{c}) \geq
d_\ml(\v{a},\v{c})$ and the condition that $d(\v{a},\v{b}) = 0$ only
if $\v{a}=\v{b}$ hold in all examples we have tried when $\v{a},
\v{b}$ and $\v{c}$ are chosen to be codewords in some code $\C$.  In
other words, it is unknown to us whether $d_\ml$ is a metric when restricted to a code
$\C \subset \{0,1\}^n$, even though it is not a metric on the entire
ambient space $\{0,1\}^n$.

\section{Appendix: Details of the simulations}\label{methods}
        
   \subsection{Generation of 1D RF codes}\label{1dimRF}
To generate the 1D RF codes used in our simulations, we took the length of the stimulus space
  to be 1, and identified the points $0$ and $1$ since the stimuli represent angles in $[0,\pi)$.
 Each receptive field (tuning curve) was chosen to be an arc of the stimulus space.  We chose our receptive fields to have a constant radius of $0.08$, which corresponds to a radius of $14.4^\circ$ in the orientation selectivity model. This parameter matches that in \cite{Somersetal95}, where tuning curves in the visual cortex were set to have half-width-half-amplitudes of $14.9^\circ$, based on experimental data from \cite{WatkinsBerkley74, Orban84}.   Each receptive field was specified by its center point.  We used 75 receptive fields to cover the stimulus space, and so our codewords had length 75.  The centers of the receptive fields were selected uniformly at random from the stimulus space, with the following modification: while the stimulus space remained uncovered, the centers were placed randomly in the uncovered region.  This modification allowed us to guarantee that the stimulus space would be covered by the receptive fields; we used a fine grid of 300 uniformly-spaced test points to find uncovered regions in the stimulus space.  
   
   By examining all pairwise intersections of receptive fields, we found all the regions cut out by the receptive fields, and each such region defined a codeword (see Fig. 1A).  Note that each codeword corresponds to a convex region of the stimulus space.  The \emph{center of mass} of a codeword is the center point of the interval to which the codeword corresponds.

   \subsection{Generation of 2D RF codes}\label{2dimRF}
To generate the 2D RF codes used in our simulations, we took the stimulus space to be a $1 \times 1$
square box environment. Each receptive field was the intersection of the stimulus space with a disk whose center lay within the stimulus space.  All disks were chosen to have the same radius; this  is consistent with findings that place fields in the dorsal hippocampus are generally circular and of similar sizes \cite{Jungetal94, Maureretal05}.  We chose the radius of our receptive fields to be $0.15$, i.e.\ $15\%$ of the width of the stimulus space, to produce codes having a reasonable sparsity of $\sim 0.07$.   As with the 1D RF codes, we generated 75 receptive fields to cover the space, with each receptive field identified by its center point.  In our simulations, the center points of the receptive fields were dropped uniformly at random in the stimulus space, with the same modification as for the 1D RF codes: while the space remained uncovered, the centers of the disks were placed uniformly at random in regions of the space that had yet to be covered.  We used a fine grid of $300 \times 300$ uniformly-spaced test points to find uncovered regions in the stimulus space.
   
   Again, by examining all intersections of receptive fields, we found all regions cut out by the receptive fields, and each region defined a codeword (see Fig. 1B). Unlike with the 1D RF codes, however, the codeword regions in the 2D RF codes were not guaranteed to be convex or even connected subsets of the stimulus space, although the typical region was at least connected.  For the purpose of defining a stimulus space distance on these codes, we defined the \emph{center of mass} of a codeword to be an appropriate approximation of the center of mass of the region corresponding to the codeword, regardless of whether that center lay within the region.  When the codeword region was large enough to contain points from the $300 \times 300$ fine grid, we took the center of mass of the codeword to be the center of mass of the grid points contained in the codeword region.  A small number of codewords had regions that were narrow crescents or other small shapes that avoided all grid points; in these cases the center of mass of the codeword was taken to be the center of mass of the receptive field boundary intersection points that defined the region.      
   
For Figure 7, we generated 10 new 2D RF codes of length 10.  For these smaller codes, the radius was chosen to be $.25$ to ensure reasonable coverage of the space.  All other parameters were as described above.
     
   \subsection{Details of error correction simulations}
      
   As a result of the chosen receptive field radii, the mean sparsity of the 1D RF codes was $s=0.165$, while the mean sparsity of the 2D RF codes was $s=0.069$.  To test how effective each of these types of codes were compared to the random codes with matched parameters, we chose to make the error probabilities as high as possible while still abiding by our BAC channel constraints and maintaining a reasonable value for the expected number of errors in each transmission.  Thus, we set $q = 0.20$ for the 1D RF codes, and $q = 0.10$ for the 2D RF codes.\footnote{In addition to the simulations shown here with the above parameters, we also tested the code performance over a range of  both larger and smaller values of $q$ and obtained similar results.}  
   To test the performance of these codes over varying degrees of channel asymmetry, the value of $p$ was chosen to range from $.05$ to $.15$ in increments of $.01$ for the 1D RF codes, while $p$ ranged from $.01$ to $.06$ in increments of $.005$ for the 2D RF codes.

\bibliographystyle{apacite}
\bibliography{neuro-coding-bibliography}

\end{document}